\documentclass[aip,reprint]{revtex4-1}
\usepackage[version=4]{mhchem}
\usepackage{textcomp}
\usepackage{upgreek}
\usepackage{graphicx}
\usepackage{multirow}
\usepackage{float}

\draft 

\newcommand{\bgo}{$\beta$-Ga$_2$O$_3\;$}
\newcommand{\go}{\ce{Ga2O3}}
\newcommand{\sio }{\ce{Si$\;$}}

\begin{document}


\title{Accumulation and removal of Si impurities on \bgo arising from ambient air exposure} 


\author{J. P. McCandless}
\email[Author to whom correspondence should be addressed: ]{jmccandless@cornell.edu}
\affiliation{School of Electrical and Computer Engineering, Cornell University, Ithaca, New York 14853, USA}

\author{C. A. Gorsak}
\email[]{Equally contributed 1st author}
\affiliation{Department of Materials Science and Engineering, Cornell University, Ithaca, New York 14853, USA}

\author{V. Protasenko}
\affiliation{School of Electrical and Computer Engineering, Cornell University, Ithaca, New York 14853, USA}

\author{D. G. Schlom}
\affiliation{Department of Materials Science and Engineering, Cornell University, Ithaca, New York 14853, USA}
\affiliation{Kavli Institute at Cornell for Nanoscale Science, Ithaca, New York 14853, USA}
\affiliation{Leibniz-Institut f\"{u}r Kristallz\"{u}chtung, Max-Born-Str. 2, Berlin, 12489, Germany}

\author{Michael O. Thompson}
\affiliation{Department of Materials Science and Engineering, Cornell University, Ithaca, New York 14853, USA}
 
\author{H. G. Xing}
\affiliation{School of Electrical and Computer Engineering, Ithaca, New York 14853, USA}
\affiliation{Department of Materials Science and Engineering, Cornell University, Ithaca, New York 14853, USA}
\affiliation{Kavli Institute at Cornell for Nanoscale Science, Ithaca, New York 14853, USA}

\author{D. Jena}
\affiliation{School of Electrical and Computer Engineering, Cornell University, Ithaca, New York 14853, USA}
\affiliation{Department of Materials Science and Engineering, Cornell University, Ithaca, New York 14853, USA}
\affiliation{Kavli Institute at Cornell for Nanoscale Science, Ithaca, New York 14853, USA}

\author{H. P. Nair}
\affiliation{Department of Materials Science and Engineering, Cornell University, Ithaca, New York 14853, USA}

\date{\today}

\begin{abstract}
Here we report that the source of Si impurities commonly observed on (010) \ce{\beta-Ga2O3} is from exposure of the surface to air. Moreover, we find that a 15 minute HF (49\%) treatment reduces the \sio density by approximately 1 order of magnitude on (010) \bgo surfaces. This reduction in Si is critical for the elimination of the often observed parasitic conducting channel, which negatively affects transport properties and lateral transistor performance. After the HF treatment the sample must be immediately put under vacuum, for the Si fully returns within 10 minutes of additional air exposure. Lastly, we demonstrate that performing a 30 minute HF (49\%) treatment on the substrate before growth has no deleterious effect on the structure or on the epitaxy surface after subsequent \go\ growth. 
\end{abstract}


\maketitle 

In recent years, there has been growing interest in the field of \go-based field-effect transistors (FETs). \go, particularly in the $\beta$-phase, has emerged as a very promising ultra-wide bandgap semiconductor with $E_g\sim 4.7$ eV.\cite{Tippins1965} This wide bandgap, along with the availability of bulk-substrates,\cite{Kuramata2016} makes \bgo an excellent candidate for high-voltage transistors for power applications.\cite{Chabak2016,Chabak2018}

To achieve high-performance FETs, precise control of the doping density and of the spatial distribution of carriers is critical. In lateral devices, for example, a parallel conducting channel at the interface between the substrate and epitaxial layers can result in high reverse-bias leakage currents.\cite{Wong2016} This is also a challenge for realizing high-electron mobility transistors (HEMTs). In HEMTs, the carrier mobility is enhanced through the creation of 2D electron gases (2DEGs).\cite{Dingle1978} By confining carriers within a well, and spatially separating this well from the chemical doping sources, impurity scattering can be reduced leading to increased mobilities. Experimentally, however, the performance of \bgo HEMTs has been limited due to large reverse leakage currents and reduced mobilities, which have been attributed to the presence of a parallel conducting channel at the epitaxy-substrate interface.\cite{Wong2016,Ahmadi2017,Kalarickal2020}

This parallel conducting channel is thought to be caused by the accumulation of \sio impurities on \go\ substrates.\cite{Bhattacharyya2023,Azizie2023,Vogt2021_adsorption,McCandless2022} This is a known issue in other compound semiconductors such as GaAs and GaN.\cite{Fu2021,Choquette1993, Cao2010} Consequently, it is imperative to investigate and understand the source, behavior, and removal of \sio impurities to reduce their impact on the electrical performance of electron devices.

\begin{figure*}
\includegraphics[scale=0.81]{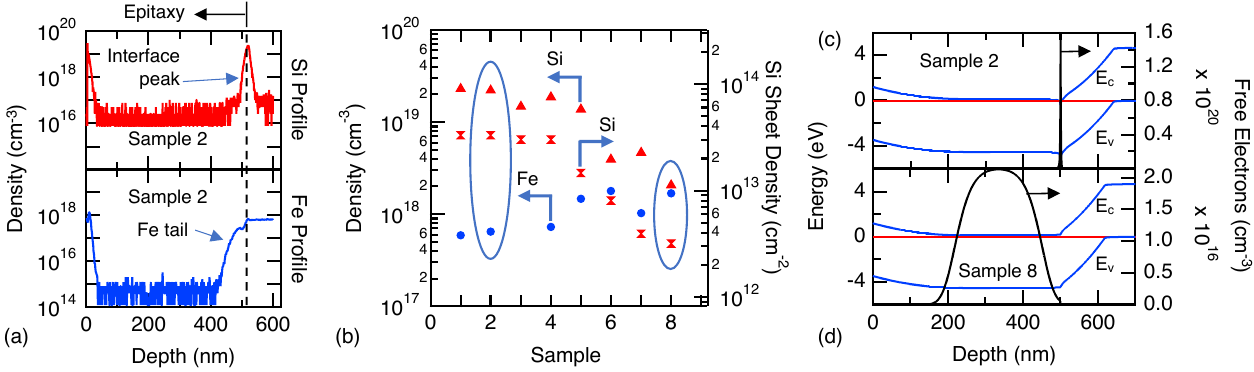}
\caption{(a) Shows SIMS measurements of a UID-\go\ sample with the Si peak at the epitaxy-substrate interface. The Fe is uniform within the substrate and then presents a tail into the film. (b) Eight samples, measured by SIMS, are shown with the average Fe density of the substrate in blue, and the interfacial Si peak value shown with red triangles. The Si interfacial peaks were integrated and the values are shown as a red hourglasses. (c) and (d) show calculated energy band diagrams with the epitaxy-substrate interface at 500 nm. (c) Shows the energy band diagram for the case where the conduction band is pulled below the Fermi level at the epitaxy-substrate due to a large density of uncompensated Si dopants. Consequently, the free carriers are confined to the interface. (d) The calculated energy band diagram for a case where the interfacial charge from \sio is fully compensated by the Fe within the Fe-\go\ substrate. As a result, the conduction band is not pulled below the Fermi level and the electron density is centered within the film.}
\label{fig:all_SIMS}
\end{figure*}

In an earlier work,\cite{McCandless2022} while addressing doping control in Si-doped \bgo grown by molecular-beam epitaxy (MBE), it was discovered that some unintentionally doped (UID) samples would show electrical conductivity while others would not when measured by Hall effect. A secondary-ion mass spectrometry (SIMS) profile of such a sample is shown in Figure~\ref{fig:all_SIMS}(a). The sample is a UID \bgo film grown by MBE on a (010) oriented edge-defined film-fed grown Fe-doped \bgo substrate from Novel Crystal Technology (NCT).\cite{Kuramata2016} At the epitaxy-substrate interface there is a large \sio peak (red trace) with a peak density of $\sim2\times10^{19}/\text{cm}^3$. The average Fe density (blue trace) within the substrate is $\sim7\times10^{17}/\text{cm}^3$. Despite being UID, with $N_d = 1.8\times10^{16}/\text{cm}^3$ in the epitaxial films, the sample displayed a free carrier sheet density of $2.2\times10^{13}/\text{cm}^2$ and a mobility of $\sim10\,\text{cm}^2/\text{V}\cdot \text{s}$ when measured by Hall effect. 

A series of other similar samples were also measured by SIMS. Figure~\ref{fig:all_SIMS}(b) shows the peak interface value of \sio (red symbols) along with the average Fe value within the substrate (blue circles) for 8 different samples. Figure~\ref{fig:all_SIMS}(b) plots the \sio peak values (red triangles, left axis) and integrated sheet values (red hour-glass, right axis). There is significant variation in \sio and Fe values among the samples. 

To understand how the Fe and Si variation affects the energy band diagram, two cases were modeled with a 1D Poisson-Schr\"{o}dinger self-consistent solver. The \sio interface peak was simulated as a sheet charge, equal to the integrated interface peak value. Figure~\ref{fig:all_SIMS}(c) shows the simulated energy band diagram for the case when there is a large difference between the \sio peak density and the Fe substrate density ($N_d = 2.22\times10^{19}/\text{cm}^3$ and $N_a = 6.49\times10^{17}/\text{cm}^3$), similar to what is observed for sample 2 in Figure \ref{fig:all_SIMS}(b). Due to the large uncompensated interfacial \sio charge, the conduction band is pulled below the Fermi level. This results in a large density of free carriers confined to the interface. Figure~\ref{fig:all_SIMS}(d) shows the simulation for the case where the \sio peak density and the Fe substrate density are of similar values ($N_d = 1.69\times10^{18}/\text{cm}^3$ and $N_a = 2.07\times10^{18}/\text{cm}^3$), like what is observed for sample 8 in Figure \ref{fig:all_SIMS}(b). In this case, the conduction band does not go below the Fermi level, and the profile of the free carriers is centered within the epitaxial film with a sheet density of $n_s = 4.6\times10^{11}/\text{cm}^2$, nearly 2 order lower than in Figure~\ref{fig:all_SIMS}(c) [$n_s = 3.1\times10^{13}/\text{cm}^2$]. Figure~\ref{fig:all_SIMS}(d) is an example of the desired condition with no parallel conduction channel. 

This \sio and Fe variation, and the corresponding variation in free carrier densities hinders the successful design and operation of FETs. To address this variation and the resulting uncompensated charges, we examined methods aimed at reducing or removing the \sio from the interface. We found that exposing the sample surface to HF (49\%) for at least 15 minutes before growth, reduced the Si density by $\sim1$ order of magnitude.

\begin{figure*}
\includegraphics[scale=0.8]{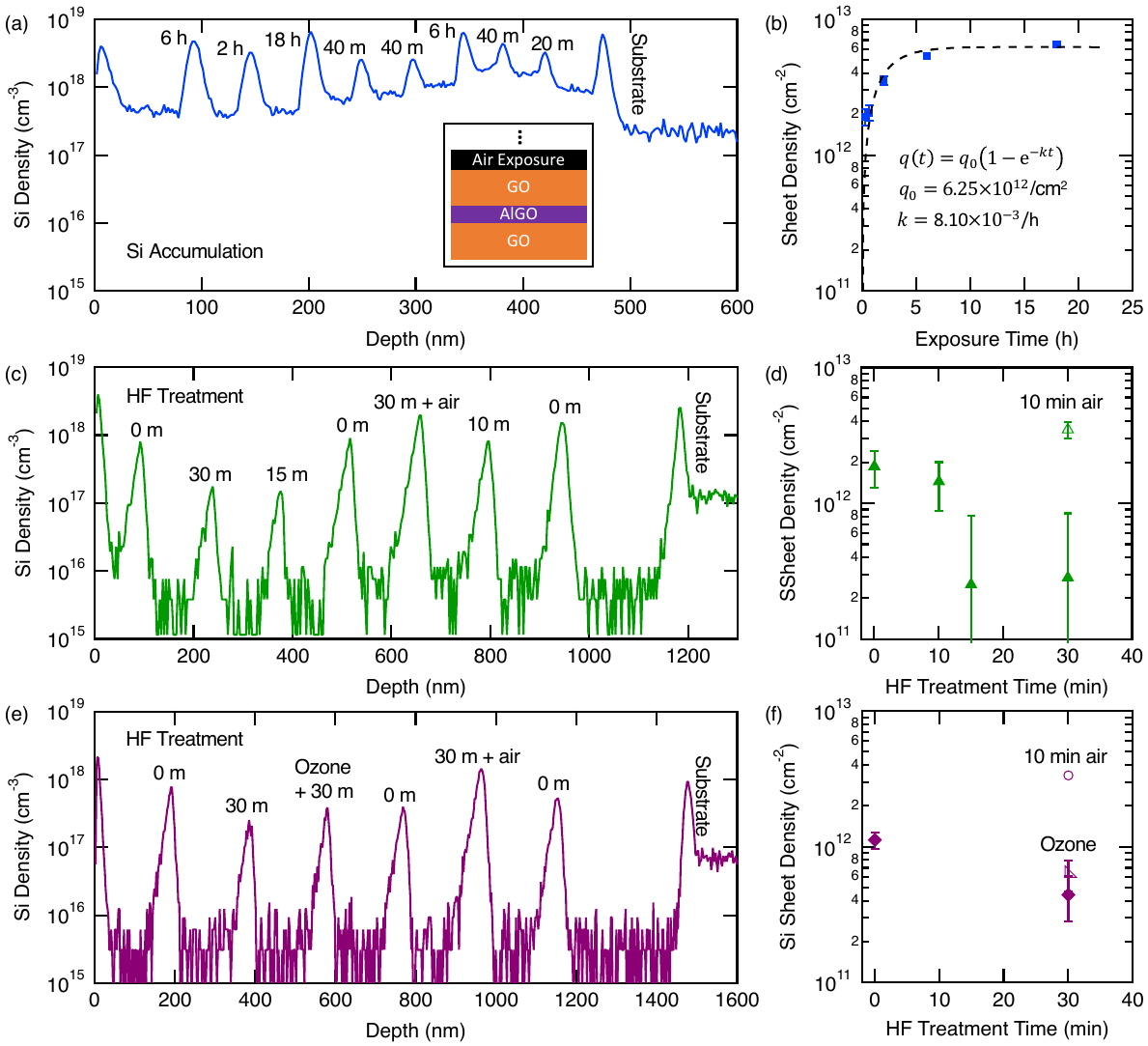}
\caption{(a)-(b) are to study the accumulation of \sio on \go\ surfaces. (a) Shows the Si profile obtained by SIMS for a UID-\go\ sample grown by MBE. Each peak corresponds to the amount of time the virgin \go\ surface was exposed to air. In between each exposure UID-\go\ is grown. Exposure times ranged from 20 minutes to 18 hours. Inset: shows the repeating layer structure for each experiment. (b) Plots the integrated sheet density obtained from panel (a) as a function of the exposure time. (c)-(f) are to study the prospect of removing \sio with HF from \go\ surfaces grown by MOCVD. (c) For each layer, the virgin surface is exposed to air for 2 hours, and then the surface is treated in HF (49\%) for the time specified above each peak. (d) Plots the integrated peak density obtained in panel (c) as a function of HF exposure time. (e) Follows the same process as described for (c) with the addition of a UV-ozone treatment step. (f) Shows the integrated peak values from panel (e).}
\label{fig:SIMS_stack}
\end{figure*}

A quantitative study of the uncompensated charge is complicated due to the significant \sio interfacial variability among substrates. To determine the efficacy of various surface treatments, it is essential to know the sheet density \emph{a priori}. There are several possible sources of the Si contaminates including siloxanes in the air,\cite{Shields1996} the colloidal silica slurry used for chemical-mechanical polishing of the substrates,\cite{Liao2023,Liu2008} quartz plasma bulbs in MBE systems or quartz chambers in metalorganic chemical vapour deposition (MOCVD) systems,\cite{Asel2020, Ikenaga20222} and re-deposition of Si from the growth chamber walls.\cite{Welser2008, Alema2021}

To investigate the accumulation of \sio from siloxanes in the air, MBE was used to grow UID-\go\ on $1\times1\,\text{cm}^2$ (010) Fe-\go\ substrates from NCT. Before loading, the substrate was sonicated in acetone, IPA, and then DI water for 5 minutes each. A Ga flux of $1.1\,\text{nm}^{-2}\text{s}^{-1}$, O flux of $2.0\,\text{nm}^{-2}\text{s}^{-1}$, and plasma power of 250 W was used during the growth. The growth temperature, based on a thermocouple reading, was 770 \textdegree C. During the growths, the MBE chamber walls are kept cold by flowing liquid nitrogen through the chamber cryoshroud to prevent possible re-deposition of \sio from the chamber walls onto the sample surface. After growing a $\sim43\,\text{nm}$ thick \go\ layer, the sample was unloaded from the MBE and moved to a fume hood where it was exposed to air for varying amounts of times, as specified in Figure~\ref{fig:SIMS_stack}(a) above each peak. Then, the sample was reloaded in the MBE for the growth of another $\sim43\,\text{nm}$ \go\ layer. In the middle of each \go\ layer, a $4.5\,\text{nm}$ \ce{(Al,Ga)2O3} (Al$\sim8\%$) layer was grown as a marker for the SIMS measurements (see layer structure in Figure~\ref{fig:SIMS_stack}(a) inset).

Figure~\ref{fig:SIMS_stack}(a) shows the Si contamination, measured by SIMS, as a function of the depth. The residence time of the sample in the fume hood between \go\ layer growths is noted above each Si peak. At this time, it is unknown why the background Si density (mid-$10^{17}$/cm$^3$) is higher than what is observed in Figure~\ref{fig:all_SIMS}(a) [2.27$\times10^{16}$/cm$^3$]. Figure~\ref{fig:SIMS_stack}(b) plots the sheet density obtained by integrating each peak in Figure~\ref{fig:SIMS_stack}(a) as a function of the air exposure time, which ranged from 20 minutes to 18 hours. Three different layers were exposed for 40 minutes from which the standard error was calculated ($\pm\,2.68\times10^{11}\,\text{cm}^3$) and was used for generating the error bars; the error bars are obscured by the symbols on the plot. The \sio sheet density saturates to $6.25\times10^{12}/\text{cm}^2$, reaching 90\% after $4.75$ hours of air exposure. The data is fit by a Lagergren pseudo-first order kinetic equation $q(t) = q_0 (1-\text{e}^{-kt})$ where $q_0 = 6.25\times10^{12}\,\text{cm}^{-2}$ and the rate constant $k = 8.10\times10^{-3}\,\text{h}^{-1}$.\cite{Lagergren1898} 

Next, we investigated the use of 49\% hydrofluoric acid (HF) treatment as a possible method of reducing the \sio surface contamination. For Figure~\ref{fig:SIMS_stack}(c)-(f),  metal-organic chemical vapour deposition (MOCVD) was used instead of MBE. The motivation for using MOCVD is that the sample can be loaded and the reactor pumped faster than in the MBE system; this minimizes the potential for additional Si accumulation during the loading process. 

As before, $1\times1\,\text{cm}^2$ substrates were solvent cleaned and then loaded into an Agnitron Agilis 100 MOCVD system. Triethylgallium (TEGa) and trimethylaluminum (TMAl) were used as gallium and aluminum precursors, respectively, and ultra-high purity molecular oxygen (99.994\%) was used as the oxidant. Ultra-high purity Ar (99.999\%) was further purified with point-of-use purifiers and was used as the carrier gas. The \go\ layers were grown at a reactor pressure of 15 Torr, substrate temperature of 800 \textdegree C with a TEGa molar flow of $77\,\mu\text{mol/min}$ and a \ce{O2}/TEGa molar ratio of 580. In the middle of each \go\ layer, a \ce{(Al,Ga)2O3} (Al$\sim4\%$) marker layer was grown at 900 \textdegree C with a reactor pressure of 50 Torr with TEGa and TMAl molar flows of $77\,\mu\text{mol/min}$ and $2.6\,\mu\text{mol/min}$, respectively. 

After the growth of each layer, the sample was removed from the reactor and placed in a fume hood for 2 hours, enabling \sio to accumulate on the surface. Next, the sample was placed in HF (49\%) for the time specified above the peaks in Figure~\ref{fig:all_SIMS}(c). Finally, the sample was placed back into the reactor for the next \go\ layer growth. This process was repeated, with freshly poured HF for each HF treatment time.

Three layers were exposed to air for 2 hours with no HF treatment, and serve as the control samples [peaks labeled ``0 m'' in Figure~\ref{fig:SIMS_stack}(c)]. From these 3 layers, the standard error was estimated to be $\pm\,5.62\times10^{11}\,\text{cm}^3$. The other layers were treated with HF (49\%) between 10 minutes and 30 minutes after exposure to air and the \sio accumulated. The peaks in Figure~\ref{fig:SIMS_stack}(c) were integrated and the values are plotted as a function of HF exposure time in Figure~\ref{fig:SIMS_stack}(d). It is inconclusive whether there was a reduction in the \sio density after the 10 minute HF treatment, but after 15 minutes there is a significant reduction in the \sio sheet density by more than $85\%$. There is no statistical difference between the 15 minute and the 30 minute HF treatment. The layer labeled ``30 m + air'' was, like the others, exposed to air within the fume hood for 2 hours, treated with HF for 30 minutes, but then left in air again for 10 minutes. Within 10 minutes, the accumulated \sio exceeded that of the control layer. Note, that for all layers there is an additional, unavoidable $\sim 5$ minute exposure to air while the sample is transported from the fume hood to the reactor, loaded, and pumped down.

The last sample, for which the SIMS profile is shown in Figure~\ref{fig:SIMS_stack}(e), was a repeat of the experiment in Figure~\ref{fig:SIMS_stack}(c), with the following exceptions: (i) for the ``30 m + air'' experiment, after the 30 minute HF treatment the sample was left on a lab bench for 10 minutes instead of in the fume hood; (ii) for the layer labeled ``Ozone + 30 m'' the sample was placed in a tabletop UV-ozone cleaner for 20 minutes after being exposed to air for 2 hours. This was in an effort to further oxidize any Si on the surface before the 30 minute HF treatment. 

The results seen in Figure~\ref{fig:SIMS_stack}(f) again show that the \sio was removed with a 30 minute HF treatment, and that the \sio re-accumulated on the \go\ surface after 10 minutes in the air. In this study, there was no clear benefit from performing an ozone treatment step. It is worth noting, however, that the sample was placed on a Si carrier wafer inside the UV-ozone cleaner, which may have affected the result.  

Lastly, for the sample shown in Figure~\ref{fig:SIMS_stack}(e), the bare substrate was cleaned by solvents, treated in HF (40\%) for 30 minutes, etched with piranha for 15 minutes, exposed to the table top UV-ozone cleaner for 20 minutes, and finally treated in HF again for 30 minutes before the growth. The sheet density of this substrate peak is $1.5\times10^{12}/\text{cm}^2$. While this is the \emph{lowest} interfacial \sio value observed in this study, it is premature to conclude whether this is due to the natural variation among the substrates, or if this reduction is, in fact, due to the surface treatment.

Based on the data reported here, and until more statistical work can be performed, we suggest the following recommendations: samples should be treated in HF (49\%) for at least 15 minutes and immediately loaded into the growth chamber. Alternatively, SIMS measurements can be performed to quantify \sio and Fe densities to ensure there is no uncompensated charge. Finally, care should be taken when studying interfacial impurities and the potential removal of the Si, since it is critical to know the Si value \emph{before} any treatment due to the significant surface variability. 

\begin{figure}
\includegraphics[scale=0.79]{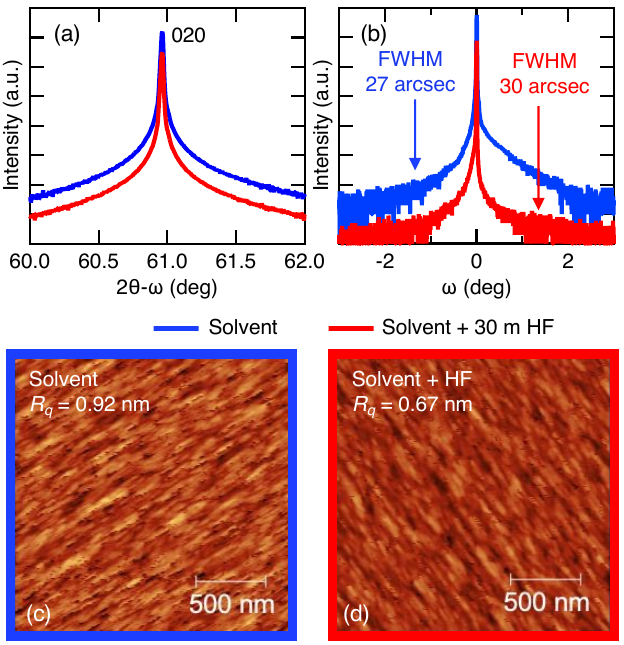}
\caption{The structural quality [(a) and (b)] and surface morphology [(c) and (d)] for 2 samples grown under the same conditions. One substrate was cleaned with solvents only (blue) and one substrate received a 30 minute HF treatment after the solvents (red). The samples were co-loaded and $\sim550$ nm of \bgo was grown. (a) Coupled $2\theta-\omega$ measurements of the 020 Bragg peak were measured. (b) Rocking curves of the 020 Bragg peak show similar crystal qualities. The surface roughnesses of the samples reveled by AFM are similar (c) and (d).}
\label{fig:growth_Compare}
\end{figure}

To understand the impact of a 30 minutes HF treatment on subsequent epitaxial growth, \bgo was grown on 2 substrates concurrently. Both substrates were first cleaned by solvents, and then one was treated in HF (49\%) for 30 minutes while the other one was not. The samples were then co-loaded and $\sim550$ nm of \bgo was grown by MOCVD. The growth conditions were the same as those described above except there was no \ce{(Al,Ga)2O3} layer, and a 100  nm low temperature \go\ buffer layer was grown first at 600 \textdegree C after which the growth temperature was increased to 800 \textdegree C for the remainder of the growth.\cite{Bhattacharyya2023}

To assess the structural quality, rocking curve measurements of the 020 diffraction peak were measured [Figure \ref{fig:growth_Compare}(b)]. The full-width at half max (FWHM) was 27 arcseconds post-growth for the sample cleaned with solvent only, while the sample with the 30 minute HF treatment in addition to the solvent clean had a FWHM of 30 arcseconds. After growth, the RMS surface roughness, measured by atomic force microscopy (AFM), revealed that the two samples have comparable roughnesses, 0.92 nm and 0.67 nm, for the solvent only and the solvent plus 30 minute HF treatment, respectively [Figure \ref{fig:growth_Compare}(c) and (d)]. This indicates that a 30 min HF treatment did not negatively impact the resulting crystalline quality or the surface morphology.

In summary, our findings reveal substantial variation in \sio and Fe concentrations within as-received (010) Fe-\ce{Ga2O3} substrates. This variation has important implications for the carrier compensation, and consequently on the electrical properties and performance of devices fabricated on a given substrate. Our investigation indicates that ambient air is a significant source of Si contamination, and that treating the substrate with HF (49\%) for a minimum of 15 minutes can reduce the Si impurity level by approximately one order of magnitude. To further reduce the potential for a parasitic conducting channel, compensation doping or \emph{in-situ} etching and removal of the interfacial Si will be required.  

\begin{acknowledgments}
This research is supported by the Air Force Research Laboratory-Cornell Center for Epitaxial Solutions (ACCESS), monitored by Dr. Ali Sayir (FA9550-18-1-0529). JPM acknowledges the support of a National Science Foundation Graduate Research Fellowship under Grant No. DGE–2139899. CAG acknowledges support from the National Defense Science and Engineering Graduate (NDSEG) Fellowship. This work uses the CCMR and CESI Shared Facilities partly sponsored by the NSF MRSEC program (DMR-1719875) and MRI DMR-1338010, and the Kavli Institute at Cornell (KIC).

\end{acknowledgments}


\bibliography{Si_accumulation.bib}

\end{document}